\PassOptionsToPackage{table}{xcolor}
\documentclass[letterpaper]{article} % DO NOT CHANGE THIS
\usepackage{aaai2027}  % DO NOT CHANGE THIS
\usepackage[hyphens]{url}  % DO NOT CHANGE THIS
\usepackage{graphicx} % DO NOT CHANGE THIS
\usepackage{natbib}  % DO NOT CHANGE THIS AND DO NOT ADD ANY OPTIONS TO IT
\usepackage{caption} % DO NOT CHANGE THIS AND DO NOT ADD ANY OPTIONS TO IT
\usepackage{algorithm}
\usepackage{algorithmic}
\usepackage{multirow}
\usepackage[table]{xcolor}
\usepackage{amssymb}
\usepackage{subcaption}
\usepackage{threeparttable}
\usepackage{newfloat}
\usepackage{listings}
\DeclareCaptionStyle{ruled}{labelfont=normalfont,labelsep=colon,strut=off} % DO NOT CHANGE THIS
\floatstyle{ruled}
\newfloat{listing}{tb}{lst}{}
\floatname{listing}{Listing}

\usepackage{booktabs}

\usepackage{amsmath}
\title{PromptTTSD: Controllable Text to Spoken Dialogue Synthesis with Open-Ended Text Descriptions}

\author{
    Sulong Yang\textsuperscript{\rm 1,\rm 2},
    Zhi Zeng\textsuperscript{\rm 1}\corresponding,
    Xiupeng Wu\textsuperscript{\rm 1},
    Juewen Hu\textsuperscript{\rm 2},\\
    Pring Wong\textsuperscript{\rm 2},
    Shuwu Zhang\textsuperscript{\rm 1}\corresponding,
    Shuo Chen\textsuperscript{\rm 2}\corresponding
}

\affiliations{
    \textsuperscript{\rm 1}Beijing University of Posts and Telecommunications\\
    \textsuperscript{\rm 2}National Key Laboratory of General Artificial Intelligence, BIGAI
}

\begin{document}

\maketitle

\begin{abstract}
Spoken dialogue synthesis aims to generate multi-turn conversational speech from dialogue scripts while maintaining speaker consistency and contextual appropriateness. However, existing models largely lack fine-grained controllability, which is critical for capturing the dynamic, context-dependent style shifts in natural conversations. While prompt-based TTS has enabled such control for single-speaker speech, existing datasets heavily rely on predefined tags, which fundamentally limits their capacity to capture open-ended styles.
To address this gap, we propose a scalable pipeline for extracting conversational speech from multi-narrator audiobooks, featuring an improved two-stage alignment algorithm to preserve dialogue continuity, paired with LLM-based extraction of style descriptions directly from narrative text. We release PromptDialog, a large-scale multi-speaker conversational speech dataset comprising 8,383 hours of Chinese speech and 120 hours of English speech, with open-ended style descriptions for 79\% of utterances. We further introduce PromptTTSD, a spoken dialogue synthesis model built via a prompt-aware fine-tuning strategy that integrates style prompts into text–speech interleaved sequences.
PromptTTSD is an open-source spoken dialogue synthesis model that supports both zero-shot voice cloning and natural language style control. Experiments demonstrate that PromptTTSD not only maintains strong fundamental synthesis capabilities but also achieves competitive open-ended style control compared with proprietary models, and substantially improves contextual appropriateness in multi-turn dialogues.

\end{abstract}

\begin{links}
    % Wrap long URLs at path boundaries.
    \def\UrlBigBreaks{}
    \def\UrlBreaks{\do\/\do\-}
    \def\UrlBreakPenalty{0}
    \renewcommand{\link}[2]{%
        \par\textbf{#1} --- %
        \pdfstartlink attr {/Border [0 0 0]} user {/Subtype /Link /A << /S /URI /URI (#2) >>}%
        \url{#2}\pdfendlink}
    \link{Demo}{https://huggingface.co/spaces/alphappp/PromptDialog-demo}
    \link{Dataset}{https://huggingface.co/datasets/alphappp/PromptDialog-v1}
    \link{Code}{https://github.com/bigai-ai/PromptTTSD}
\end{links}

\section{Introduction}
Spoken dialogue synthesis aims to generate natural multi-speaker conversations from written dialogue scripts while maintaining speaker consistency and contextual appropriateness. Recent systems have made substantial progress in long-form generation, speech quality and speaker consistency~\citep{peng2025vibevoice,xie2025soulx}. However, existing systems largely lack fine-grained controllability. Such control is particularly important in dialogue synthesis, where speaking behavior depends not only on textual content but also on the conversational context. An inappropriate speaking style can disrupt the emotional coherence of an interaction. For example, a cheerful delivery in a conversation about bereavement would be incongruous with the underlying scenario.

\begin{figure*}[t]
    \centering
    \includegraphics[width=\textwidth]{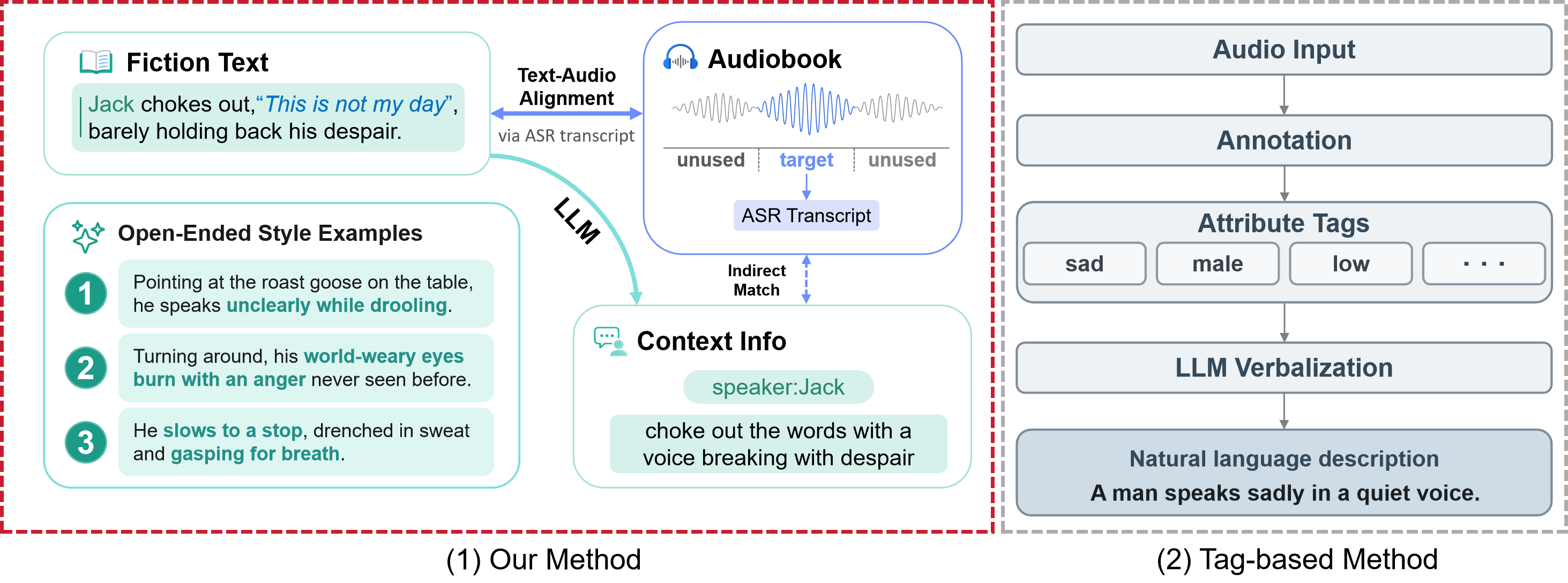}
    \caption{Comparison of pipelines for constructing speech datasets with natural-language descriptions. In (1), the open-ended examples are drawn from our dataset, and the bold text highlights open-ended style elements that are difficult for external tools to annotate.
}
    \label{fig:intro}
\end{figure*}

Natural language control offers a promising solution to this limitation. Existing prompt-based TTS systems have demonstrated that speaking behavior can be controlled through textual descriptions, but such control has been explored primarily in single-speaker synthesis~\citep{du2024cosyvoice,hu2026qwen3,guo2023prompttts}. Moreover, descriptions in existing expressive speech datasets are typically constructed from predefined tags obtained through external tools or human annotation. These tags are then verbalized into descriptions by large language models (LLMs), as illustrated in Fig.~\ref{fig:intro} (2). Consequently, existing methods remain closed-ended, as their descriptions are restricted to predefined speaking styles. They therefore fail to capture open-ended styles arising from facial expressions, physical actions, compound emotions, and special speaker states, as illustrated by the open-ended examples in Fig.~\ref{fig:intro} (1). Such nuanced behaviors are difficult to identify reliably using external tools.

% To address this bottleneck, we turn to  multi-narrator audiobooks as a naturally rich yet underutilized source of conversational speech. We find that voice acting in these audiobooks is highly expressive and closely matches the descriptions associated with the corresponding dialogue in the source book. As illustrated in the red dashed box of Fig.~\ref{fig:intro}, we use an LLM to derive descriptions and speaker identity from the narrative context surrounding each dialogue utterance. We then use Automatic Speech Recognition (ASR) transcripts to locate the audiobook segments corresponding to the extracted dialogue text, yielding conversational speech paired with transcripts and natural language descriptions.

To address this bottleneck, we first build a dialogue synthesis dataset by using multi-narrator audiobooks as a naturally rich yet underutilized source of conversational speech.
Specifically, our key insight is that voice acting in these audiobooks is highly expressive and closely matches the open-ended descriptions associated with the corresponding dialogue in the source books. To leverage this resource, the audiobook recordings must be aligned with original books, so that each speech segment can be paired with its transcript and narrative context.
LibriHeavy~\citep{kang2024libriheavy} provides a strong foundation for such audio–-text alignment. However, reliably aligning long-form audiobook recordings with their original books at utterance level without disrupting the continuity of multi-turn dialogues remains a severe challenge. To overcome this, we propose a robust, context-aware alignment algorithm that effectively recovers incomplete utterance boundaries and preserves dialogue continuity.

Based on the aligned audio--text pair, we further use an LLM to extract descriptions and speaker identities from the matched source text, as illustrated in Fig.~\ref{fig:intro} (1). Using this pipeline, we construct \textbf{PromptDialog}, a large-scale multi-speaker conversational speech dataset, comprising 8,383 hours of Chinese speech and 120 hours of English speech. Every utterance is paired with a speaker identity, and 79\% of the utterances are annotated with a textual description.

Furthermore, to bridge the gap between open-ended textual descriptions and expressive dialogue generation, we introduce \textbf{PromptTTSD}. Rather than relying on simple tag embeddings, we propose a description-aware tuning framework that effectively integrates natural language descriptions into interleaved text–speech representations. We further evaluate open-ended style control and description-conditioned conversational speech synthesis. Results show that PromptTTSD achieves comparable performance on standard TTS metrics. For open-ended style control, it is competitive with proprietary models and outperforms existing open-source systems. For conversational speech synthesis, conditioning on descriptions leads to better contextual appropriateness.

Our main contributions are summarized as follows:

\begin{itemize}
\item We propose the pipeline for constructing natural-language-annotated conversational speech datasets, built on a robust and efficient alignment algorithm that produces naturally aligned text–speech–description triplets from audiobooks and their source fiction texts.
\item We release PromptDialog, a large-scale multi-speaker conversational speech dataset annotated with natural language descriptions, comprising 8,383 hours of Chinese speech and 120 hours of English speech.
\item We introduce PromptTTSD, an open-source spoken dialogue synthesis model that supports natural language control. PromptTTSD performs competitively with proprietary models in evaluations of open-ended control. Separately, in dialogue synthesis, conditioning on descriptions substantially improves contextual appropriateness.

\end{itemize}

\section{Related Work}

\subsection{Expressive Text-to-Speech}

Expressive TTS can be broadly categorized into three paradigms: control via discrete tags, control via reference speech, and control via natural language instructions. \citet{lee2017emotional} leverage emotion tags to control the affective quality of synthesized speech, while \citet{guo2023emodiff} further introduce intensity control on top of emotion tags. \citet{li2021controllable} achieve emotion cloning through reference audio. However, both tag-based and reference-based control lack flexibility and expressiveness. To address this limitation, several studies have explored natural-language control~\citep{guo2023prompttts,leng2023prompttts,shimizu2024prompttts++,yang2024instructtts}. \citet{du2024cosyvoice,du2024cosyvoice2,du2025cosyvoice3} and \citet{hu2026qwen3} adopt LLM-based architectures and large-scale speech corpora, substantially advancing natural language control over speaking styles. However, their datasets are not publicly released, making their data construction and training pipelines difficult to reproduce.

\subsection{Multi-Speaker Conversational Speech Synthesis}

\citet{nguyen2023generative} were the first to employ a dual-tower Transformer for text-free spoken dialogue synthesis. \citet{zhang2024covomix} achieve spoken dialogue synthesis with natural overlapping turns. \citet{ju2025mooncast} were the first to adopt an LLM as the backbone of a spoken dialogue synthesis model that supports up to four speakers. \citet{peng2025vibevoice} introduce a continuous speech tokenizer, supporting long-form audio synthesis of up to 90 minutes. \citet{xie2025soulx} incorporate dialect and paralinguistic tags to generate more natural and personalized podcast speech. Despite this progress, existing multi-speaker spoken dialogue synthesis systems have largely been limited to tag-level control, whereas utterance-level control through natural language descriptions remains underexplored. This limits their ability to specify diverse, context-dependent speaking styles in conversational speech.

\subsection{Natural Language Stylistic Datasets}

\begin{table*}[t]
\centering

\begin{tabular}{lcccccc}
\toprule
\multirow{2}{*}{\textbf{Dataset}}
& \multicolumn{3}{c}{\textbf{Dataset Scale}}
& \multicolumn{3}{c}{\textbf{Dataset Properties}} \\
\cmidrule(lr){2-4}
\cmidrule(lr){5-7}
& \textbf{Duration}
& \textbf{Clips}
& \textbf{Speakers}
& \textbf{Speech Type}
& \textbf{Description Source}
& \textbf{Open Source} \\
\midrule

NLSpeech~\citep{yang2024instructtts}
& 44h
& --
& 7
& Monologue
& 2 labels
& $\times$ \\

PromptSpeech~\citep{guo2023prompttts}
& --
& 28k
& --
& Monologue
& 5 labels
& $Partial_1$ \\

PromptTTS 2~\citep{leng2023prompttts}
& --
& 20k
& --
& Monologue
& 4 labels
& $\times$ \\

TextrolSpeech~\citep{ji2024textrolspeech}
& 330h
& 236k
& 1,324
& Monologue
& 5 labels
& $\checkmark$ \\

SpeechCraft~\citep{jin2024speechcraft}
& 2,391h
& 2,250k
& $>$3,200
& Monologue
& 8 labels
& $\checkmark$ \\

SpeechCaps~\citep{huang2025speechcaps}
& --
& --
& 4
& Dialogue
& 5 labels
& $Partial_2$ \\

DNASpeech~\citep{cheng2025dnaspeech}
& 18h
& 22k
& 2,395
& Monologue
& Movie scripts
& $\checkmark$ \\

\rowcolor{gray!20}
PromptDialog (ours)
& \textbf{8,503 (6,859) h}
& \textbf{4,769k}
& $>$1,000
& Dialogue
& Fiction texts
& $\checkmark$ \\

\bottomrule
\end{tabular}

\caption{ Comparison of stylistic speech datasets. 
For PromptDialog, the value in parentheses indicates the duration of utterances with non-empty descriptions.
\textit{Partial}$_1$ indicates that approximately 15\% of the corpus is publicly released. \textit{Partial}$_2$ indicates that only the text is publicly released, while the audio is not publicly available. }
\label{tab:stylistic_speech_datasets}

\end{table*}

\citet{guo2023prompttts,leng2023prompttts,shimizu2024prompttts++,guan2024mm,yang2024instructtts,ji2024textrolspeech,jin2024speechcraft} derive predefined attributes of speech through manual annotation or automatic recognition. These attributes are then verbalized into prompts with LLMs. Despite differences in annotation strategy and granularity, the semantic scope of each description is constrained by a predefined schema. Such descriptions therefore provide limited coverage of attributes beyond the selected inventory. \citet{huang2025speechcaps} extend this paradigm to multi-speaker speech by combining synthesized single-speaker speech into conversations. Leveraging movie scripts, \citet{cheng2025dnaspeech} obtained both natural language descriptions and dialogue context, with the descriptions moving beyond conventional tag-based annotation schemes. However, the dialogue context is provided only in textual form, so the dataset cannot serve as a multi-speaker conversational speech dataset. Comparison between stylistic speech datasets is shown in Tab.~\ref{tab:stylistic_speech_datasets}.

\begin{figure}
    \centering
    \includegraphics[width=\linewidth]{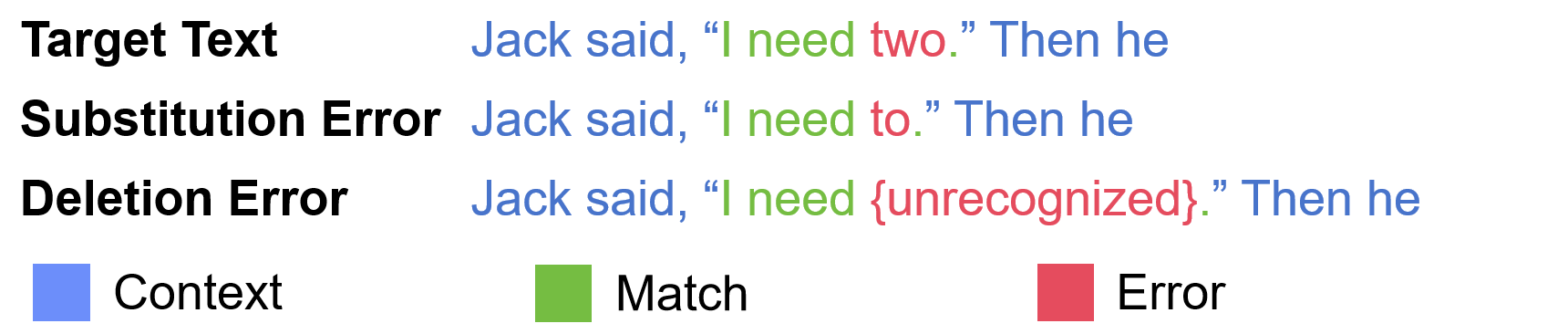}
    \caption{Examples of partial speech alignment errors produced by the LibriHeavy alignment pipeline.}
    \label{fig:alignment}
\end{figure}

\section{Dataset}
\subsection{Data Collection}

We focus on multi-narrator audiobooks, where different characters in dialogue can be distinguished by their voices without narration. Unlike conventional read-aloud audiobooks, these productions resemble television dramas without the visual component, combining narration with acted dialogue and rich vocal expression.

We searched YouTube for audiobooks that met these criteria and obtained the corresponding books from legitimate online sources. In total, we collected 107 Chinese audiobooks and 20 English audiobooks, yielding 8,383 hours of Chinese dialogue speech and 120 hours of English dialogue speech. English audiobooks are generally shorter than Chinese audiobooks, so collecting English data required substantially more manual effort. We plan to expand the English portion of the dataset in future work.

\subsection{Audio Alignment}

LibriHeavy provides an efficient pipeline for aligning
audiobooks with their original books. However, as illustrated in Fig.~\ref{fig:alignment}, LibriHeavy often fails to align utterance prefixes or suffixes affected by deletion or substitution errors, even when the remaining content can be correctly matched. Such partial matches can be removed by filtering when constructing independent speech segments, but removing even a single utterance can reduce the continuity and semantic coherence of a multi-turn dialogue. We therefore propose a two-stage alignment method that incorporates contextual boundary expansion and directly aligns automatic transcripts with full-length book texts without text segmentation.

\paragraph{Text processing.}
For automatic speech recognition (ASR), we use Whisper-large-v3~\citep{radford2023robust} for English audiobooks and Paraformer-zh~\citep{gao2023funasr} for Chinese audiobooks. We normalize the original fiction texts before subsequent processing. Specifically, for English, we follow the text-normalization procedure adopted by LibriHeavy. For Chinese, we first apply the FunASR text normalizer and then convert the normalized text into tone-stripped Pinyin tokens. The latter representation reduces mismatches caused by homophonic substitutions in the automatic transcripts.

\paragraph{Stage 1: Anchor-based alignment.}
Let \(Q=(q_1,\ldots,q_M)\) denote the full token sequence of the
original book, and let \(T=(t_1,\ldots,t_N)\) denote the full token
sequence of the timestamped ASR transcript for the corresponding
audiobook. Following the close-match strategy of LibriHeavy, we
construct a suffix array over \(T\) and use quoted-text spans
\(Q_{i:j}=(q_i,\ldots,q_j)\) as queries to retrieve candidate
locations in the target transcript.

A verified alignment is represented by an anchor
\(a=(i,j,m,n)\), indicating that the quoted-text span \(Q_{i:j}\)
is aligned with the ASR span \(T_{m:n}=(t_m,\ldots,t_n)\).

The anchors are maintained in monotonic order. For two adjacent
anchors,
\[
\begin{aligned}
a_k &= (i_k,j_k,m_k,n_k),\\
a_{k+1} &= (i_{k+1},j_{k+1},m_{k+1},n_{k+1}),
\end{aligned}
\]
we require
\[
\begin{aligned}
j_k &< i_{k+1},\\
n_k &< m_{k+1}.
\end{aligned}
\]
A query span located between these anchors is searched only within
the corresponding interval of the target transcript,
\[
T_{n_k+1:m_{k+1}-1}.
\]

At initialization, two anchors are placed at the boundaries of the query and target sequences:
\[
\begin{gathered}
a_0 = (0,0,0,0),\\
a_{K+1} = (M+1,M+1,N+1,N+1).
\end{gathered}
\]
where \(K\) is the number of verified anchors. 
Consequently, the first candidate search covers the entire target transcript, whereas
subsequent searches are progressively restricted by newly verified
anchors. This allows the quoted dialogue to be aligned directly with
the full-length automatic transcript without prior segmentation of
the transcript.

Candidate query spans are processed in descending order of length.
Longer spans typically yield fewer candidate locations that satisfy the matching criterion, thereby reducing the number of candidates that need to be verified.

For each candidate alignment
\(\bigl(Q_{i:j},T_{m:n}\bigr)\), where \(Q_{i:j}\) is a
quoted-text query span and \(T_{m:n}\) is a retrieved candidate span
in the ASR transcript, we compute their token-level Levenshtein distance~\citep{levenshtein1966binary},
denoted by
\(d_{\mathrm{lev}}\bigl(Q_{i:j},T_{m:n}\bigr)\).

Let
\[
\ell=\max\bigl(|Q_{i:j}|,|T_{m:n}|\bigr)
\]
denote the maximum length of the two spans. The candidate alignment
is accepted if
\[
d_{\mathrm{lev}}\bigl(Q_{i:j},T_{m:n}\bigr)
\leq \tau(\ell),
\]
where
\[
\tau(\ell)=
\begin{cases}
0, & \ell\leq 9,\\[3pt]
\left\lfloor\ell/5\right\rfloor, & \ell>9.
\end{cases}
\]

An accepted candidate alignment is inserted into the anchor set,
thereby restricting the search intervals for subsequent shorter
query spans. Once the anchors are fixed, the search intervals bounded
by different pairs of adjacent anchors are independent, allowing
candidate retrieval and verification to be parallelized across
multiple CPU cores.

\paragraph{Stage 2: Context-based alignment.}
The first stage identifies reliable matching spans. However, an
accepted span may have incomplete boundaries when the prefix or
suffix of the ASR span contains deletion or
substitution errors.

Given a candidate pair \((C_Q,C_T)\), we compute their longest common
subsequence (LCS), the longest sequence of tokens shared by both spans
in the same order, without requiring contiguity. We denote the
corresponding matched token positions by
\[
\mathcal{M}
=
\{(r_1,z_1),\ldots,(r_H,z_H)\},
\]
where \(H\) is the number of matched token pairs, and
\(r_h\) and \(z_h\) are relative token indices in \(C_Q\) and
\(C_T\), respectively, satisfying
\[
\begin{aligned}
r_1<\cdots<r_H,\\
z_1<\cdots<z_H.
\end{aligned}
\]
We additionally calculate the length ratio
\[
\rho=
\frac{\min(|C_Q|,|C_T|)}
     {\max(|C_Q|,|C_T|)}.
\]
If both boundaries of the candidate are supported by corresponding
tokens in the other sequence and \(\rho>0.7\), the candidate is
regarded as complete.

Otherwise, we estimate the amounts of unmatched content at the boundaries from the first and last LCS matches:
\[
\delta_{\mathrm{L}}
=
\max(r_1-1,z_1-1),
\]
\[
\delta_{\mathrm{R}}
=
\max(|C_Q|-r_H,\;|C_T|-z_H).
\]
We then construct left and right context windows with sizes
\[
\begin{aligned}
w_{\mathrm{L}}=2\delta_{\mathrm{L}}, \\
w_{\mathrm{R}}=2\delta_{\mathrm{R}},
\end{aligned}
\]
and search for additional correspondences in the surrounding query
and target contexts. Once reliable contextual matches are found,
the candidate boundaries are extended to the outermost matched
context positions.

The contextual matches serve as boundary evidence rather than
requiring every token within the expanded span to be recognized
correctly. Consequently, content in the quoted dialogue that is
omitted or incorrectly transcribed by ASR can still be retained
within the aligned query span when it lies between reliable
contextual matches. This is particularly useful for preserving
paralinguistic vocalizations, such as laughter and sighs, as well
as acoustically similar words that are incorrectly recognized by
ASR.

\subsection{Dialogue Grouping and Description Extraction}
\label{sec:description_extraction}

We group quoted utterances into dialogue groups according to their
positions in the source text. Specifically, two consecutive quoted
utterances are assigned to the same group if they are separated by
no more than two intervening lines.

For each utterance, we define its description as a nearby narrative
cue that directly characterizes or informs how the utterance is
spoken. Such cues may describe emotion, tone, gesture, action, facial expression, or other contextual information. In the prompts provided to the LLMs, we instruct them to preserve the wording of the source text whenever possible and to avoid unnecessary summarization or paraphrasing. If no valid descriptive cue is available, the description is set to an empty string.

Each dialogue group is represented as a JSON array, where each object contains a quoted utterance and a corresponding description field. Based on the speaker attribution results reported in Table~\ref{tab:speaker_attribution}, and considering the inference cost of annotation, we select MiMo-v2.5~\citep{xiaomi2026mimo25} and prompt it to complete the description field.

We subsequently perform consistency checks on the model
output. Specifically, we verify that the numbers of input and output items are identical and that each output utterance corresponds exactly to its input utterance. These checks prevent malformed responses and unintended modifications to the source dialogue text from entering the dataset.

\subsection{Speaker Attribution}
\label{sec:speaker_attribution}

We obtain speaker identities directly from the source fiction using
an LLM. Since speaker attribution and description extraction rely on the same dialogue context, we perform the two tasks jointly in a
single model call. Specifically, for each quoted utterance, the model is required to complete both the \texttt{speaker} and
\texttt{description} fields while leaving the dialogue text
unchanged. 

Conventional pipelines~\citep{zhu2026zipvoice,xie2025soulx} for constructing multi-speaker conversational speech datasets typically rely on voice activity detection (VAD) to segment speech and speaker diarization to assign speaker labels. However, acoustic speaker diarization remains challenging when the number and identities of speakers are not known in advance. A recent benchmark of state-of-the-art diarization systems reports
that even the lowest overall diarization error rate among the
evaluated systems remains \(11.2\%\), achieved by
PyannoteAI\footnote{\texttt{https://www.pyannote.ai/}}, with missed
speech and speaker confusion constituting the primary sources of
error~\citep{lanzendorfer2025benchmarking}.

We therefore leverage the narrative context available in the original book and prompt LLMs to infer the speaker of each quoted utterance. As shown in Table~\ref{tab:speaker_attribution}, speaker information derived from the book text yields substantially more accurate speaker labels than the acoustic baseline, which relies on VAD and acoustic speaker clustering.

We further verify the assigned speaker labels using speaker embeddings
extracted with CAM++~\citep{wang2023cam++}. Utterances with low
similarity to their assigned speaker are compared with other speakers.
When an alternative speaker meets the similarity threshold, we provide
the similarity scores to Qwen3.8-27B~\citep{qwen2026qwen38} to determine whether the label should
be corrected. The verification thresholds and additional refinement
steps are detailed in the ``Speaker Refinement'' section of the
supplementary material.

\section{Experiments}

\subsection{Dataset Evaluation}
\subsubsection{Alignment Quality.}

\begin{table}[t]
\centering
\small
\begin{tabular}{lcc}
\toprule
Method & Sentence Match (\%) $\uparrow$ & Word Match (\%) $\uparrow$ \\
\midrule
LibriHeavy & 71.39 & 89.12 \\
Ours & \textbf{94.13} & \textbf{94.38} \\
\bottomrule
\end{tabular}
\caption{Comparison of audio--text alignment methods on a manually annotated evaluation set.}
\label{tab:alignment}
\end{table}

We evaluate alignment quality on 10 hours of audiobook data. For each quoted utterance in the reference text, we use both methods to extract the corresponding audio segment and transcribe it using the same ASR model. For quoted utterances containing more than five words, an extracted segment is automatically considered correct if its word error rate is below 5\% and its first and last words match those of the reference utterance. Short utterances and samples that do not meet these criteria are manually verified. We report sentence match rate and word match rate as evaluation metrics.

As shown in Table~\ref{tab:alignment}, our method improves the sentence match rate from 71.39\% to 94.13\%, while also increasing the word match rate from 89.12\% to 94.38\%. The larger improvement in sentence match rate indicates that our method more reliably recovers complete quoted utterances, whereas LibriHeavy often aligns only part of an utterance despite matching most of its words.

\subsubsection{Description Quality.}
We randomly sampled 500 utterances for human evaluation. A description was considered valid if it was non-empty, stylistically informative, and consistent with the perceived speech. Generic descriptions such as ``said'' were treated as invalid. Overall, 62\% of the sampled utterances contained valid descriptions. No substantive description contradicted the speech; invalid cases were mainly missing or overly generic. We report this figure as a reference estimate, since open-ended speech descriptions cannot yet be evaluated reliably at scale.

\subsubsection{Diversity of Style Descriptions.}

To provide a more interpretable comparison of
the expressive styles captured by different
datasets, we conduct a semantic analysis of
PromptDialog and SpeechCraft using an open
vocabulary of style tags. Specifically, we use
Qwen3.8-27B~\citep{qwen2026qwen38} to extract tags from both datasets
following a unified protocol. For PromptDialog,
we first filter narrative contexts to retain
information relevant to utterance delivery,
while dialogue text is used only to refine
attributes supported by the context. We then
group semantically related tags using text
embeddings and clustering refined by an LLM,
excluding groups containing fewer than three
tag entries. The resulting tags are organized
into seven categories for a detailed comparison
of expressive diversity. Further details are
provided in the supplementary material.

\begin{table}[t]
\centering
\small
\setlength{\tabcolsep}{4pt}
\begin{tabular}{lcccc}
\toprule
Category & P & S & P-only & S-only \\
\midrule
Facial expressions & 714 & 123 & 597 & 6 \\
Vocal delivery & 1,343 & 970 & 774 & 401 \\
Emotion/attitude & 1,253 & 1,123 & 385 & 255 \\
Cognitive/verbal & 662 & 315 & 380 & 33 \\
Physiological states & 238 & 108 & 141 & 11 \\
Mixed states & 131 & 31 & 108 & 8 \\
Scene interactions & 283 & 18 & 270 & 5 \\
\midrule
\textbf{Total} & \textbf{4,624} & \textbf{2,688}
& \textbf{2,655} & \textbf{719} \\
\bottomrule
\end{tabular}
\caption{Semantic tag diversity on equally sized samples
(1,083,438 records per dataset). P and S denote PromptDialog
and SpeechCraft. P-only and S-only count tag types observed
exclusively in the corresponding dataset after semantic
consolidation.}
\label{tab:description_semantic_diversity}
\end{table}

To control for corpus size, we compare equally sized samples
of 1,083,438 records from each dataset.
As shown in Table~\ref{tab:description_semantic_diversity},
PromptDialog covers 4,624 retained tag types, versus 2,688
in SpeechCraft (72.0\% more), with 2,655 and 719
dataset-exclusive types, respectively.
The advantage is evident in vocal delivery (1,343 vs.~970)
and cognitive/verbal activities (662 vs.~315), and is
particularly pronounced for facial expressions (714 vs.~123).
These results demonstrate that PromptDialog descriptions cover
a broader range of expressive and contextual cues.

\subsubsection{Speaker Attribution Quality.}
To evaluate the reliability of LLM-based speaker attribution, we sample 100 dialogue segments from our dataset and compare it with a commercial speaker diarization system. We use pyannoteAI Precision-2~\citep{pyannoteai2025precision2} as the speaker diarization baseline. We evaluate both methods only on dialogue regions, excluding narration. Both methods are evaluated using word-level speaker attribution error against the same manually annotated ground truth. For LLM-based attribution, the predicted speaker label is assigned to every word in that utterance. For speaker diarization, the predicted speaker label of each word is obtained from the diarization output. The error rate is the proportion of words assigned to an incorrect speaker.

As shown in Table~\ref{tab:speaker_attribution}, LLM-based speaker attribution substantially outperforms speaker diarization, suggesting that inferring speakers from fiction text is more reliable than relying solely on acoustic information for speaker diarization.

\begin{table}[t]
\centering
\small
\setlength{\tabcolsep}{4pt}
\begin{tabular}{lc}
\toprule
Method & Error Rate (\%) $\downarrow$ \\
\midrule
\multicolumn{2}{l}{\textit{Speaker Diarization}} \\
pyannoteAI Precision-2~\citep{pyannoteai2025precision2}
& 8.70 \\
\midrule
\multicolumn{2}{l}{\textit{LLM-based Speaker Attribution}} \\
GPT-5~\citep{openai2025gpt5}
& \textbf{0.29} \\
% GPT-5-mini~\citep{openai2025gpt5}
% & 0.48 \\
Gemini 2.5 Pro~\citep{googledeepmind2025gemini25pro}
& 0.62 \\
% Qwen3.5 Plus~\citep{qwen2026qwen35}
% & 0.63 \\
DeepSeek-v4-Pro~\citep{deepseek2026v4}
& \underline{0.41} \\
% GLM-5.1~\citep{zai2026glm51}
% & 0.48 \\
MiMo-v2.5~\citep{xiaomi2026mimo25}
& 0.50 \\
\bottomrule
\end{tabular}
\caption{Comparison of speaker diarization and LLM-based speaker attribution on target dialogue regions.}
\label{tab:speaker_attribution}
\end{table}

\subsection{Training Setup}
% \subsubsection{Base Model and Trainable Components}
% \subsubsection{Training Objective}
% \subsubsection{Input Construction}
% \subsubsection{Implementation Details}
We adopt SoulX-Podcast~\citep{xie2025soulx}, which does not natively support natural language control, as the base model for controllable spoken dialogue synthesis. Since its downstream conditional flow-matching model and vocoder achieve high-quality speech reconstruction, we freeze these components during training. We train only the LLM backbone to learn the relationship between the textual prompt and the target speech. The remaining training details are described below.

\paragraph{Token Organization.}
Following SoulX-Podcast, we retain the text--speech interleaved sequence representation for sentence-by-sentence dialogue synthesis. In this formulation, the text tokens of each utterance are followed by the corresponding speech tokens, and utterances from different speakers are concatenated in temporal order. Each utterance begins with a speaker token indicating the speaker identity.

Unlike SoulX-Podcast, which inserts a dialect token after the speaker token, we replace the dialect token with a sequence of natural language prompt tokens that describe the intended speaking style. The resulting dialogue sequence is organized as follows:
\[
\begin{aligned}
&\langle \mathrm{SPEAKER1}\rangle
\langle \mathrm{Prompt}\rangle
\langle \mathrm{Text}\rangle
\langle \mathrm{Audio}\rangle \\
&\langle \mathrm{SPEAKER2}\rangle
\langle \mathrm{Prompt}\rangle
\langle \mathrm{Text}\rangle
\langle \mathrm{Audio}\rangle \\
&\langle \mathrm{SPEAKER3}\rangle \cdots
\end{aligned}
\]

\paragraph{Learning Objective.}
To encourage the LLM backbone to capture the semantic information conveyed by the style descriptions, we introduce an auxiliary text prediction loss in addition to the original speech-token prediction objective. The overall training objective is defined as
\begin{equation}
    \mathcal{L}_{\mathrm{total}}
    =
    \mathcal{L}_{\mathrm{speech}}
    +
    \lambda \mathcal{L}_{\mathrm{text}},
\end{equation}
where $\mathcal{L}_{\mathrm{speech}}$ denotes the speech loss, $\mathcal{L}_{\mathrm{text}}$ denotes the auxiliary text loss, and $\lambda$ is set to $0.01$. We examine this weight and the prompt placement in the Ablation subsection.

\paragraph{Training Configuration.}
We fine-tune the model for 6 epochs on 15 PPU-ZW810E GPUs using dynamic batching. The model is optimized with AdamW. The learning rate follows a warmup followed by cosine decay schedule, with a peak learning rate of $5\times10^{-5}$ and 2,500 warmup steps. Training is performed for 36k optimization steps in total.

\begin{table*}[t]
\centering
\setlength{\tabcolsep}{4pt}
\begin{tabular*}{\textwidth}{
    @{\extracolsep{\fill}}
    lccccc
    @{}
}
\toprule
\multirow{2}{*}{Model}
& \multicolumn{2}{c}{Subjective Metrics}
& \multicolumn{3}{c}{Objective Metrics} \\
\cmidrule(lr){2-3}
\cmidrule(lr){4-6}
& Style-CMOS $\uparrow$
& Q-MOS $\uparrow$
& UTMOS $\uparrow$
& SIM $\uparrow$
& CER $\downarrow$ \\
\midrule

CosyVoice3~\citep{du2025cosyvoice3}
& $-0.273 \pm 1.326$
& $4.173 \pm 0.947$
& $3.164$
& $\mathbf{0.646}$
& $4.897$ \\

MiMo-v2.5-TTS~\citep{xiaomi2026mimov25tts}
& $-0.993 \pm 1.114$
& $3.760 \pm 0.939$
& $3.179$
& $0.588$
& $\mathbf{3.973}$ \\

\textbf{PromptTTSD} (seed-tts voice)
& $\mathbf{0.000}$
& $\mathbf{4.520 \pm 0.631}$
& $\mathbf{3.436}$
& $0.626$
& $5.039$ \\

\midrule

Eleven-v3~\citep{elevenlabs2025elevenv3}
& $\mathbf{0.007 \pm 1.298}$
& $4.600 \pm 0.733$
& $3.247$
& $0.421$
& $8.555$ \\

\textbf{PromptTTSD} (eleven voice)
& $0.000$
& $\mathbf{4.750 \pm 0.591}$
& $\mathbf{3.390}$
& $\mathbf{0.662}$
& $\mathbf{5.515}$ \\

\bottomrule
\end{tabular*}
\caption{
Utterance-level open-ended style control results.
Subjective metrics are reported as mean $\pm$ standard deviation
computed over all individual ratings.
}
\label{tab:utterance_style_results}
\end{table*}

\begin{table}[t]
\centering
\small
\setlength{\tabcolsep}{4pt}
\begin{tabular}{lcccc}
\toprule
Model
& CER $\downarrow$
& ACC $\uparrow$
& SIM $\uparrow$
& CA-MOS $\uparrow$ \\
\midrule

MOSS-TTSD
& $4.85$
& $0.959$
& $0.795$
& $2.86 \pm 0.85$ \\

SoulX-Podcast
& $4.29$
& $\mathbf{0.973}$
& $\mathbf{0.828}$
& $2.93 \pm 0.87$ \\

PromptTTSD
& $\mathbf{3.76}$
& $0.969$
& $0.805$
& $3.06 \pm 1.07$ \\

PromptTTSD (Desc.)
& $4.30$
& $0.968$
& $0.803$
& $\mathbf{3.52 \pm 0.95}$ \\

\bottomrule
\end{tabular}
\caption{
Evaluation results on TTSD-ZH. MOSS-TTSD is proposed by ~\citet{zhang2026moss}.
CA-MOS is reported as mean $\pm$ standard deviation
computed over all individual ratings. 
}
\label{tab:ttsd_zh_results}
\end{table}

\subsection{Model Evaluation Setup}
We evaluate PromptTTSD from two complementary perspectives:
style controllability and dialogue synthesis. The controllability evaluation compares PromptTTSD with existing prompt-based TTS systems in terms of their ability to follow natural language descriptions. The dialogue synthesis
evaluation reports standard synthesis metrics and examines whether explicit
style descriptions improve contextual appropriateness. Since our training corpus primarily consists of Chinese speech, all evaluations are conducted in Chinese.

\paragraph{Style Controllability Evaluation.}
We focus on open-ended style control, which evaluates the model's
ability to follow complex natural-language style descriptions that are
difficult to assess using predefined labels.

\textbf{Evaluation Data.}
We manually construct 50 concise style descriptions, each paired with
corresponding speech content. For each test case, we additionally annotate the
expected perceptual characteristics of the generated speech to provide
consistent guidelines for human evaluation.

\textbf{Evaluation Metrics.}
We report both subjective and objective metrics. Five native speakers
participate in the subjective listening tests. Human-perceived speech quality
is evaluated using Q-MOS, while relative adherence to the given style
description is assessed using Style-CMOS. For Style-CMOS, human evaluators compare two anonymized samples presented in random order. They rate which sample better matches the target description on a scale from $-2$ to $+2$. A score of $+2$ indicates a strong preference for the first sample, $-2$ indicates a strong preference for the second, and $0$ indicates no preference. We set PromptTTSD to $0.00$ and report each baseline relative to it. Positive scores favor the baseline, while negative scores favor PromptTTSD. We further report UTMOS~\citep{saeki2022utmos} as an automatic estimate of speech quality.
Speaker similarity (SIM) and character error rate (CER) are computed using
the official evaluation models released with
Seed-TTS-eval~\citep{anastassiou2024seed}.

\textbf{Baselines.}
We compare PromptTTSD with CosyVoice3,
MiMo-v2.5-TTS~\citep{xiaomi2026mimov25tts}, and
Eleven-v3~\citep{elevenlabs2025elevenv3}.
CosyVoice3 serves as a representative open-source baseline that supports both
zero-shot voice cloning and natural-language style control. MiMo-v2.5-TTS and Eleven-v3 are the
latest proprietary TTS offerings from Xiaomi and ElevenLabs, respectively,
and provide expressive speech generation controlled through textual style instructions or audio tags.

For CosyVoice3 and MiMo-v2.5-TTS, all systems use the same reference speech
selected from Seed-TTS-eval. For the comparison with Eleven-v3, we use voices
provided by ElevenLabs and condition PromptTTSD on the corresponding voices
to maintain a consistent speaker condition.

\paragraph{Dialogue Synthesis Evaluation.}
We evaluate spoken dialogue synthesis on the TTSD-ZH
benchmark~\citep{zhang2026moss}, following its official evaluation protocol.
We report character error rate (CER) for content accuracy, speaker attribution
accuracy (ACC) for speaker assignment, and speaker similarity (SIM) for
speaker consistency.

Beyond these standard metrics, we investigate whether explicit
utterance-level natural-language style descriptions improve the contextual
appropriateness of synthesized dialogue. TTSD-ZH does not provide style
annotations. Therefore, for each utterance, GPT-5 generates a style
description based on the utterance itself and its surrounding dialogue
context. These descriptions are used only as conditioning inputs for
PromptTTSD (Desc.), while the original PromptTTSD synthesizes speech
without descriptions.

Human evaluators are presented with the original dialogue context and the
corresponding synthesized utterance, and rate on a five-point scale how
appropriately the speech matches the conversational scenario. We report the mean rating as \textbf{Contextual Appropriateness MOS (CA-MOS)}. For example, the synthesized speech
should reflect an energetic and engaging delivery in a game-commentary
context, or an appropriately sorrowful tone in a farewell scene. By comparing
PromptTTSD with and without the generated style descriptions, we assess
whether explicit style guidance improves the contextual appropriateness of
dialogue synthesis.

\subsection{Results}
\paragraph{Style Controllability Results.}
Table~\ref{tab:utterance_style_results} reports the results of open-ended
style control. The results demonstrate that PromptTTSD provides competitive open-ended style
controllability compared with both open-source and proprietary TTS systems. PromptTTSD consistently achieves the highest Q-MOS and UTMOS scores under
both speaker conditions, indicating superior human-perceived and automatically
estimated speech quality. It also maintains strong speaker consistency during
expressive style generation. 

CER is relatively high across all evaluated systems. Our analysis suggests that this is primarily because the evaluation set contains many non-verbal vocalizations, such as laughter and sighs. When these expressions are rendered as vocal sounds, they are often incorrectly transcribed by the ASR model, resulting in a higher CER~\citep{bai2026synparaspeech}. In contrast, unnatural read-aloud realizations of the corresponding textual markers are more readily recognized by ASR and may therefore achieve a lower CER. Eleven-v3 exhibits a higher CER because it occasionally treats
parts of the style description as target text and synthesizes them as speech.

\paragraph{Dialogue Synthesis Results.}
Table~\ref{tab:ttsd_zh_results} shows that PromptTTSD without style
descriptions achieves the lowest CER among all evaluated systems and
outperforms the baselines in contextual appropriateness, while maintaining
competitive speaker attribution accuracy and speaker similarity. When
conditioned on utterance-level style descriptions,
PromptTTSD (Desc.) retains competitive CER, ACC, and SIM, but substantially
improves CA-MOS from $3.06$ to $3.52$. This result indicates that explicit
style descriptions enable PromptTTSD to generate speech that better matches
the intended dialogue context without substantially compromising content
accuracy, speaker attribution, or speaker consistency.

\subsection{Ablation}
We examine the auxiliary text loss weight and the placement of style
prompts. All ablation models are trained on a 10\% subset of the
training data for four epochs.

\paragraph{Auxiliary Text Loss Weight.}
We vary $\lambda$ from $0$ to $1$ in the training objective.
As shown in Table~\ref{tab:ablation_text_loss}, setting $\lambda=0.01$
reduces the text loss from $15.2979$ to $4.1304$, while achieving the
lowest speech loss and highest speech token accuracy among the tested
settings. Increasing the weight further reduces the text loss, with
the lowest value at $\lambda=0.5$, while speech loss increases and
speech token accuracy remains close to the baseline or decreases.
The backbone is initialized from Qwen3-1.7B, which already provides
text representations learned during pretraining. We therefore use
$\lambda=0.01$ to introduce auxiliary text supervision with a small
weight while maintaining speech prediction performance.

\begin{table}[t]
\centering
\small
\setlength{\tabcolsep}{4pt}
\caption{Auxiliary text loss weight ablation after four epochs on
10\% of the training data. Token accuracy refers to speech tokens.}
\label{tab:ablation_text_loss}
\begin{tabular}{cccc}
\toprule
$\lambda$ & Speech loss $\downarrow$ & Token acc. (\%) $\uparrow$ & Text loss $\downarrow$ \\
\midrule
0 & 2.7665 & 36.4169 & 15.2979 \\
0.01 & \textbf{2.7654} & \textbf{36.4396} & 4.1304 \\
0.05 & 2.7676 & 36.4211 & 3.7535 \\
0.1 & 2.7683 & 36.4289 & 3.6666 \\
0.5 & 2.7749 & 36.3771 & \textbf{3.6166} \\
1 & 2.7835 & 36.2953 & 3.6181 \\
\bottomrule
\end{tabular}
\end{table}

\paragraph{Prompt Placement.}
Following the practice of placing style prompts before the target
text in single speaker expressive TTS, our dialogue synthesis model
uses the same ordering. We compare this layout with placing the prompt
after the target text:
\[
\begin{aligned}
&\langle\mathrm{Speaker}\rangle
\langle\mathrm{Prompt}\rangle
\langle\mathrm{Text}\rangle
\langle\mathrm{Audio}\rangle,\\
&\langle\mathrm{Speaker}\rangle
\langle\mathrm{Text}\rangle
\langle\mathrm{Prompt}\rangle
\langle\mathrm{Audio}\rangle.
\end{aligned}
\]
We evaluate both layouts on 1,000 synthesized utterances.
As shown in Table~\ref{tab:ablation_prompt_placement}, placing the prompt
before the text yields a lower CER ($3.046\%$ vs.~$3.394\%$) and a
slightly higher speaker similarity ($0.6744$ vs.~$0.6731$).
We retain this ordering in the final model.

\begin{table}[t]
\centering
\small
\setlength{\tabcolsep}{4pt}
\caption{Prompt placement ablation on 1,000 synthesized utterances.
Before and After indicate the prompt position relative to the target text.}
\label{tab:ablation_prompt_placement}
\begin{tabular}{lcc}
\toprule
Prompt position & CER (\%) $\downarrow$ & SIM $\uparrow$ \\
\midrule
Before & \textbf{3.046} & \textbf{0.6744} \\
After & 3.394 & 0.6731 \\
\bottomrule
\end{tabular}
\end{table}

\section{Conclusion}
We presented PromptDialog, a large-scale dataset for prompt-based controllable spoken dialogue synthesis constructed from multi-narrator audiobooks and their original books. Our pipeline enables scalable extraction of speech--text--description triplets without manual annotation. Building on this dataset, we introduced a reproducible post-training framework that equips an existing spoken dialogue synthesis model with utterance-level natural language control. Experimental results show that the resulting PromptTTSD maintains strong synthesis quality and speaker consistency, achieves competitive open-ended style controllability compared with proprietary systems, and improves the contextual appropriateness of synthesized multi-turn dialogue when style descriptions are provided.

\clearpage
\appendix
\begin{center}
    {\LARGE\bfseries Appendix}
\end{center}

\section{Additional Evaluation on TTSD-Eval}

Table~\ref{tab:ttsd_eval} reports additional results on the Chinese
and English subsets of TTSD-eval. Compared with the original
SoulX-Podcast model, PromptTTSD achieves a lower Chinese CER, reducing
it from 4.29\% to 3.76\%. However, its speaker attribution accuracy and
speaker similarity decrease slightly from 0.9730 to 0.9690 and from
0.8280 to 0.8050, respectively.

One possible explanation is that continued fine-tuning changes the
speaker distribution learned by the base model. Although PromptDialog
is large in duration, the speaker coverage of the data used for
fine-tuning is more limited than that of the original pre-training
corpus. Moreover, PromptDialog consists primarily of audiobook speech,
whereas TTSD-eval contains out-of-domain conversational samples.
Fine-tuning on this relatively homogeneous domain may therefore reduce
the generalization of speaker representations to unseen voices and
recording conditions.

In contrast, the improvement in Chinese CER may benefit from the
text--audio alignment strategy used to construct PromptDialog. Unlike
pipelines that directly use ASR transcripts as training targets, our
method aligns speech with the corresponding source fiction text. The
resulting transcripts may contain fewer recognition errors and provide
cleaner supervision for speech-content modeling. This explanation is
consistent with the improved CER, although a controlled comparison
between source-text and ASR-transcript supervision would be required
to verify the effect.

For English, PromptTTSD shows slightly lower ACC and SIM and a higher
WER than SoulX-Podcast. PromptDialog contains only 120 hours of English
speech, substantially less than its Chinese portion. The limited
English fine-tuning data, together with the audiobook-domain mismatch,
may be insufficient to preserve the English generalization ability of
the base model. Nevertheless, PromptTTSD remains competitive with the
other evaluated open-source systems across all three English metrics.

\begin{table*}[t]
    \centering
    \caption{TTSD-eval results. ACC denotes speaker attribution
    accuracy, SIM represents speaker similarity, and CER and WER
    measure content accuracy for Chinese and English, respectively.
    The best and second-best results are shown in bold and underlined,
    respectively.}
    \label{tab:ttsd_eval}

    \begin{tabular*}{0.8\textwidth}{
        @{\extracolsep{\fill}}lcccccc@{}
    }
        \toprule
        \multirow{2}{*}{Model}
        & \multicolumn{3}{c}{ZH}
        & \multicolumn{3}{c}{EN} \\
        \cmidrule(lr){2-4}
        \cmidrule(lr){5-7}
        & ACC $\uparrow$
        & SIM $\uparrow$
        & CER $\downarrow$
        & ACC $\uparrow$
        & SIM $\uparrow$
        & WER $\downarrow$ \\
        \midrule

        Higgs Audio V2~\citep{higgsaudio2025}
        & -- & -- & --
        & 0.9025 & 0.6860 & 21.31\% \\

        FireRedTTS-2~\citep{xie2025fireredtts}
        & 0.9022 & 0.7383 & 7.68\%
        & -- & -- & -- \\

        VibeVoice 1.5B~\citep{peng2025vibevoice}
        & 0.8798 & 0.7415 & 8.18\%
        & 0.9353 & 0.6961 & 11.33\% \\

        VibeVoice 7B~\citep{peng2025vibevoice}
        & 0.9222 & 0.7590 & 5.70\%
        & 0.9554 & 0.7140 & 9.46\% \\

        SoulX-Podcast~\citep{xie2025soulx}
        & \textbf{0.9730}
        & \textbf{0.8280}
        & \underline{4.29\%}
        & \textbf{0.9745}
        & \textbf{0.7487}
        & \textbf{7.22\%} \\

        \textbf{PromptTTSD (ours)}
        & \underline{0.9690}
        & \underline{0.8050}
        & \textbf{3.76\%}
        & \underline{0.9699}
        & \underline{0.7276}
        & \underline{7.55\%} \\

        \bottomrule
    \end{tabular*}
\end{table*}

\section{Dialogue Context and Style Control}

\paragraph{Evaluation Setup.}
We investigate the benefit of retaining dialogue history over
generating each turn independently, and the effect of explicit style
instructions under both generation settings. We construct 50 dialogues
containing 300 utterances with pronounced emotional intent, such as
angry arguments, and annotate each utterance with a target style
description. We compare PromptTTSD, which retains dialogue
history, with a variant that generates each turn independently without
dialogue history. Both variants are evaluated with and without style
instructions. We also evaluate SoulX-Podcast without instructions.

For each synthesized utterance, Gemini Flash 3.8 makes a binary
judgment of whether the speech matches the annotated style description.
The judge receives only the current synthesized utterance and its
target description, without the surrounding dialogue context.
We report style accuracy as the percentage of utterances judged to
match their descriptions. The same target descriptions are used to
evaluate speech generated with and without instructions.

\paragraph{Results.}
As shown in Table~\ref{tab:style_accuracy}, PromptTTSD improves
style accuracy over SoulX-Podcast without instructions and also
supports explicit style control.
The comparison with independent turn generation shows that retaining
longer dialogue context improves adaptation to the intended scenario.
Explicit instructions further steer speech toward the specified
style, with PromptTTSD achieving a style accuracy of $85.7\%$.

\begin{table}[t]
\centering
\small
\setlength{\tabcolsep}{4pt}
\caption{Style accuracy (\%) on 50 dialogues containing 300 utterances,
judged by Gemini Flash 3.8. Independent turns denotes the PromptTTSD
variant that generates each turn without dialogue history.}
\label{tab:style_accuracy}
\begin{tabular}{lcc}
\toprule
Model & \shortstack{Without\\instructions $\uparrow$}
& \shortstack{With\\instructions $\uparrow$} \\
\midrule
SoulX-Podcast & 55.7 & -- \\
PromptTTSD & \textbf{76.3} & \textbf{85.7} \\
\quad Independent turns & 68.7 & 83.3 \\
\bottomrule
\end{tabular}
\end{table}

\section{Additional Details of Dataset Construction}

\subsection{Speaker Refinement}
\label{app:speaker_refinement}
For utterances whose speaker identity cannot be determined from the text, the LLM assigns the label \texttt{unknown}. We complement these annotations with acoustic speaker representations extracted using CAM++~\citep{wang2023cam++}. For each \texttt{unknown} utterance, we compute the cosine similarity between its embedding and the embeddings of known speakers appearing in the local context, and assign the utterance to the speaker with the highest similarity. We empirically set the similarity threshold to \(0.4\) based on inspection during data construction. This threshold provides a practical balance between retaining same-speaker utterances with substantial expressive variation and rejecting low-confidence matches. Utterances whose maximum similarity falls below this threshold are discarded.

We also use CAM++ speaker embeddings to verify utterances with an
assigned speaker label. An utterance is flagged for further inspection
only when its cosine similarity to the assigned speaker is below
\(0.3\). We then compare its embedding with those of other speakers.
If at least one alternative speaker has a similarity of \(0.5\) or
higher, we provide the similarity scores for the assigned and candidate
speakers to Qwen3.8-27B~\citep{qwen2026qwen38}, which determines whether the speaker label should
be corrected. Thus, the similarity thresholds select cases for review,
while the final correction decision is made by the LLM.

Speaker labels extracted from fiction are not always unique character names. Generic role labels, such as \textit{policeman} or
\textit{boy}, may refer to different characters in different local
contexts and therefore correspond to different voices. We therefore perform a second round of speaker assignment for potentially ambiguous speaker labels. Their acoustic
embeddings and neighboring speaker assignments are jointly examined
to determine whether occurrences sharing the same textual label
should be assigned to different speaker identities.

\subsection{Temporal Boundary Refinement}

Word-level timestamps produced by ASR systems are not always accurate.
Directly segmenting audio according to these timestamps may therefore
truncate the final word of an utterance. To reduce such boundary
errors, we refine the start and end timestamps using the neighboring
ASR words, while limiting the adjustment on each side to
\(\Delta_{\max}=1.0\) second.

Consider a matched word span from word \(i\) to word \(j\). Let
\(s_k\) and \(e_k\) denote the start and end timestamps of word \(k\),
respectively. If a word exists immediately before the matched span,
the refined start boundary is defined as

\begin{equation}
\tilde{s}
=
\max\left(
s_i-\Delta_{\max},
\frac{e_{i-1}+s_i}{2}
\right).
\end{equation}

Similarly, if a word exists immediately after the matched span, the
refined end boundary is defined as

\begin{equation}
\tilde{e}
=
\min\left(
e_j+\Delta_{\max},
\frac{e_j+s_{j+1}}{2}
\right).
\end{equation}

In other words, each boundary is moved toward the midpoint of the
silence or interval between the matched span and its neighboring
word, subject to a maximum adjustment of one second. If the matched
span begins with the first word or ends with the last word of the
ASR sequence, the corresponding boundary is left unchanged.

\section{Additional Dataset Statistics}

\begin{figure*}[t]
    \centering
    \includegraphics[width=\linewidth]{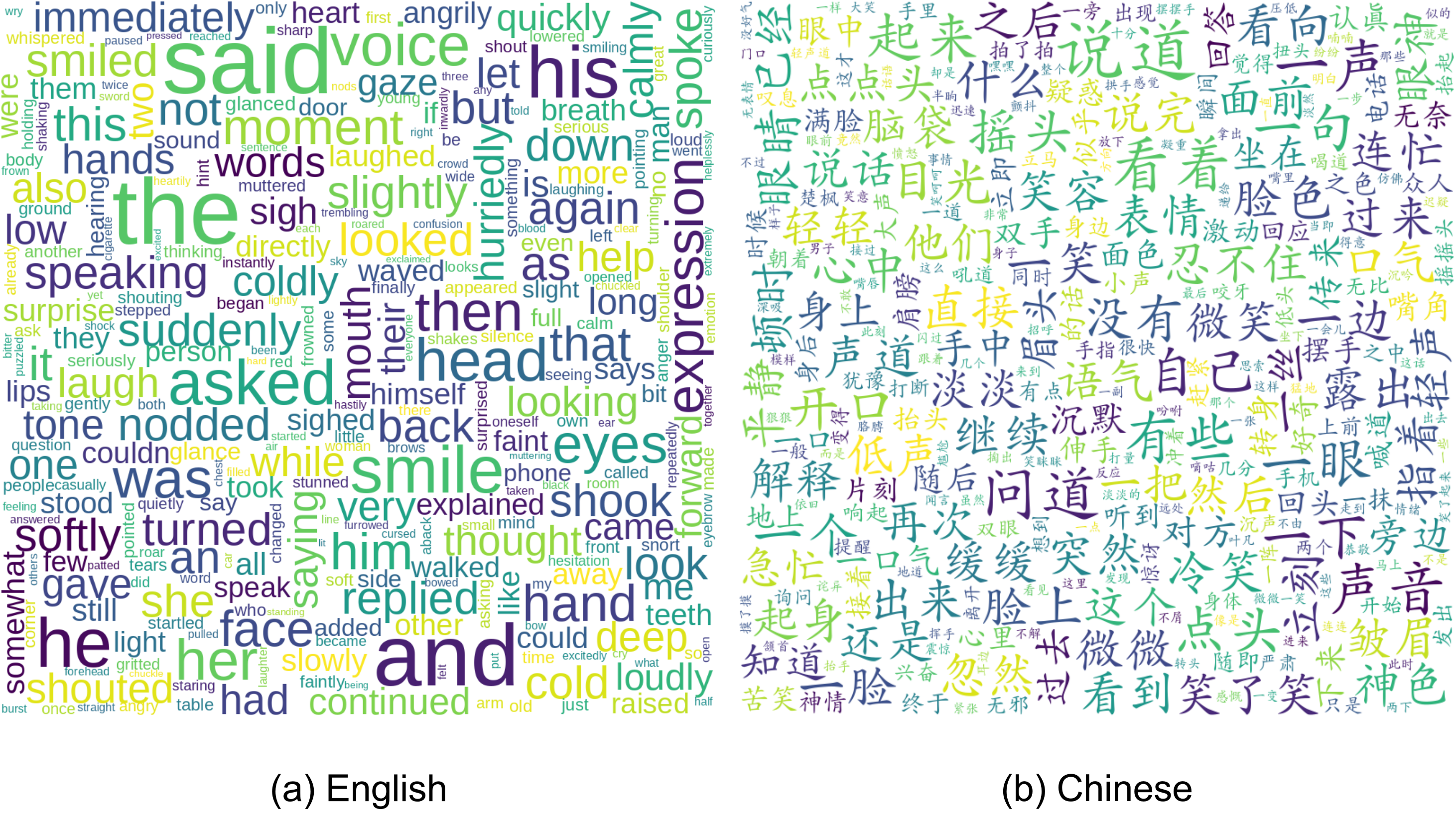}
    \caption{Word clouds of the 300 most frequent words in the
    English and Chinese descriptions of PromptDialog.}
    \label{fig:wordclouds}
\end{figure*}

\begin{figure*}[t]
    \centering
    \includegraphics[width=\linewidth]{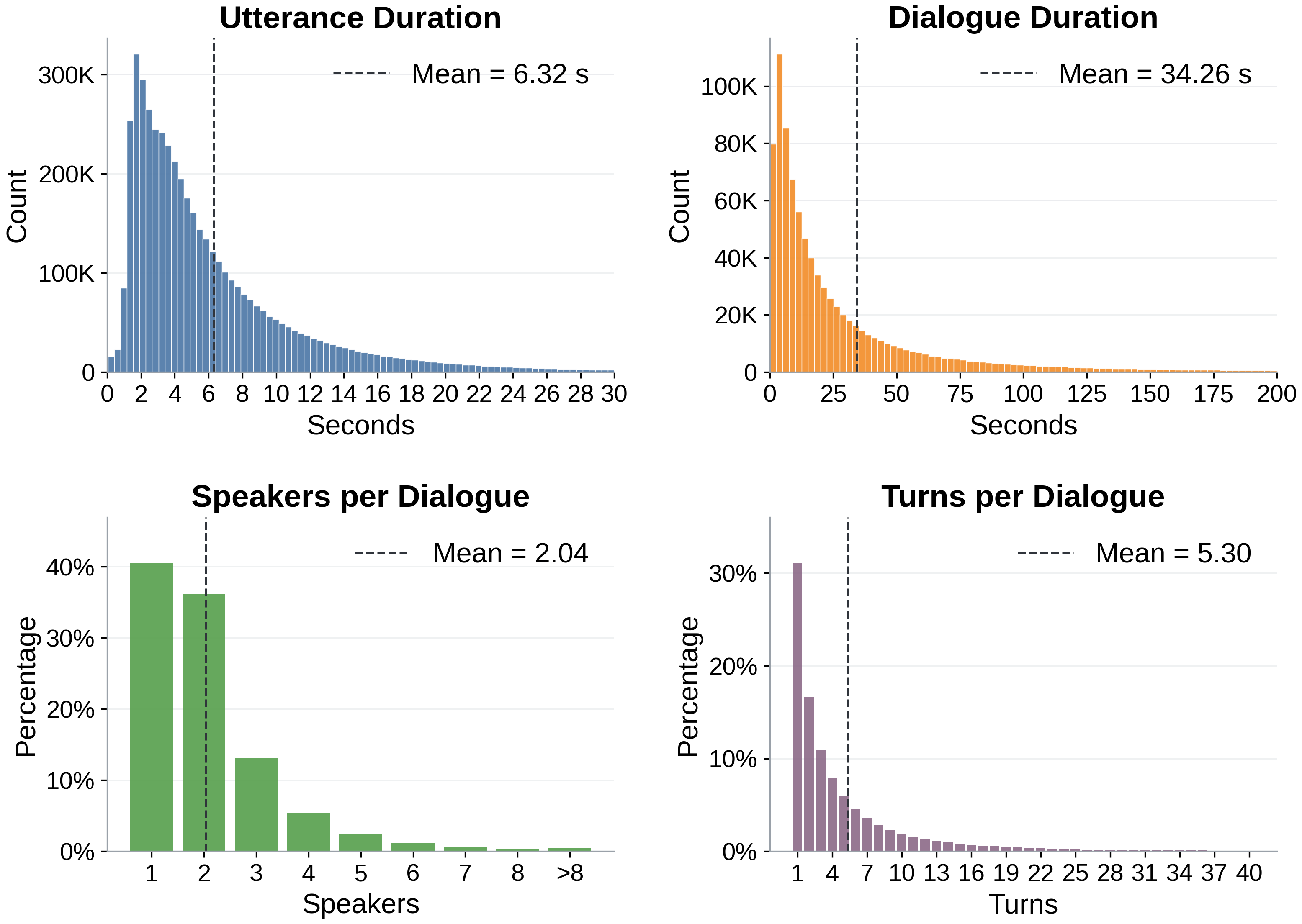}
    \caption{Distributions of utterance duration, dialogue duration,
    speaker count, and turn count in PromptDialog.}
    \label{fig:duration_distribution}
\end{figure*}

Figure~\ref{fig:wordclouds} shows the 300 most frequent words
in the English and Chinese descriptions of PromptDialog.
Figure~\ref{fig:duration_distribution} presents the distributions
of utterance duration, dialogue duration, speaker count, and turn
count.

\subsection{Additional Details on Description Diversity}
\label{app:description_diversity}

\subsubsection{Open-Ended Tag Extraction and Consolidation.}
We process PromptDialog and SpeechCraft with a common
open-ended tag extraction protocol using Qwen3.8-27B~\citep{qwen2026qwen38}.
For PromptDialog, the narrative context is first filtered
to retain only spans directly relevant to utterance delivery,
including tone, emotion, facial expression, gesture, and
voice-related or physiological states. Unrelated plot details,
temporal expressions, and actions without a direct relation
to speaking are discarded. The filtering stage preserves
source wording rather than paraphrasing or expanding it.

The tag extractor takes the filtered context and the dialogue
utterance as inputs. Candidate tag examples are illustrative
rather than exhaustive: the model may generate tags outside
the example vocabulary. Context-supported descriptions are
prioritized and preserved in their original wording where
possible. Composite cues may also be decomposed into
constituent attributes. The utterance itself can only refine
states supported by the context and must not be used to
infer new emotions or speech acts from textual content alone.
The extractor emits at most five tags per utterance, each
of at most five Chinese characters, and returns an empty
list when no supported cue is available. Equivalent criteria
are applied to SpeechCraft descriptions.

We first form preliminary semantic groups using text
embeddings and remove outliers. We then use the LLM to
inspect whether groups should be split, followed by semantic
reclustering. Finally, groups containing fewer than three
tag entries are excluded. Retained tag types are organized
into seven expressive or context-related categories.
% TODO: Include exact prompts and clustering hyperparameters
% for reproducibility, preferably as additional tables/listings.
Table~\ref{tab:exclusive_semantic_groups} lists the top-10 semantic
groups exclusive to each dataset.

\begin{table}[t]
\centering
\includegraphics[width=\linewidth]{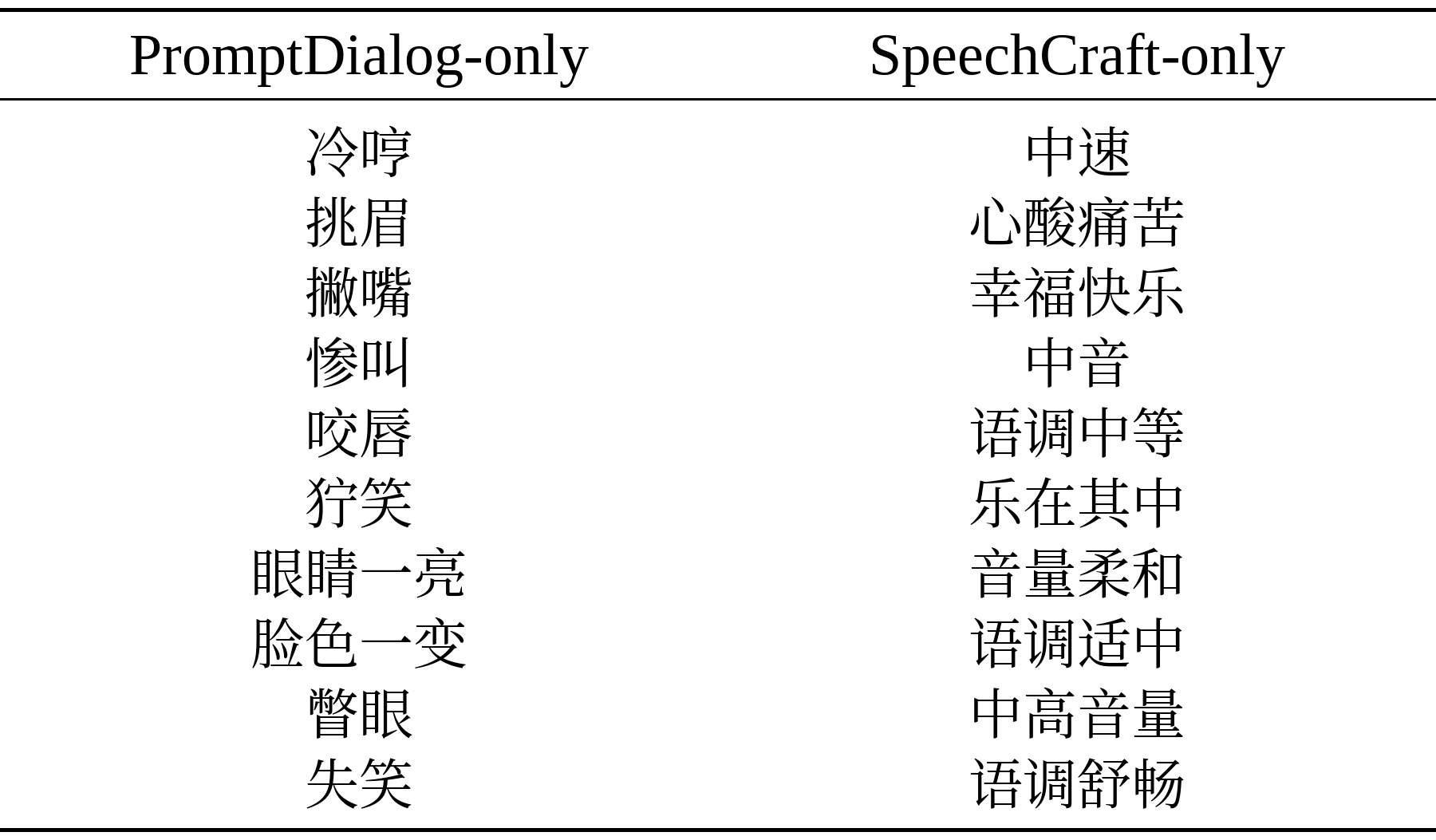}
\caption{Top-10 semantic groups exclusive to each dataset,
shown with their original Chinese labels.}
\label{tab:exclusive_semantic_groups}
\end{table}

\section{Potential Applications of PromptDialog}

Beyond controllable dialogue synthesis, PromptDialog may support
several related research directions. The applications discussed below
are preliminary and have not yet been systematically evaluated. We
present them as potential extensions enabled by the dataset and its
construction pipeline, rather than as demonstrated capabilities.

\subsection{Speech Captioning}

PromptDialog may provide additional supervision for speech captioning.
Compared with datasets constructed from a predefined attribute
inventory, its open-ended descriptions cover a broader range of
speaking behaviors, including sarcastic delivery, breathlessness, and
states associated with physical actions. Such annotations may help
extend speech captioning beyond conventional attributes such as
emotion, pitch, speaking rate, and volume.

However, descriptions extracted from fiction do not always correspond
to information that can be uniquely inferred from speech. Some
descriptions concern the surrounding scene, a character's actions, or
events that are only weakly reflected in the acoustic signal. These
descriptions may be acceptable as conditioning inputs for controllable
TTS, where the model can learn to use acoustically relevant information
and ignore unrelated contextual details. For speech captioning,
however, treating all such descriptions as prediction targets would
introduce substantial ambiguity.

For example, a fearful utterance may be described as being produced
in a frightening environment, as reflecting the speaker's emotional
state, or as a reaction to a preceding event. These interpretations
may be equally compatible with the same acoustic observation.
Consequently, directly recovering the original narrative description
from speech is generally an ill-posed objective.

A possible direction is therefore to combine speech with additional
information from the source fiction, such as dialogue context,
character identity, and narrative events. Another possibility is to
separate acoustically observable attributes from context-dependent
interpretations, rather than requiring a model to reproduce the
original description verbatim. Such formulations may provide more
appropriate supervision for open-ended speech understanding.

\subsection{Voice Design}

The proposed pipeline may also facilitate the construction of data for
text-based voice design. DeepDubbing~\citep{dai2026deepdubbing} has
explored the use of voice descriptions derived from audiobook
characters, although the details of its data acquisition and
annotation procedure are not fully specified.

Because PromptDialog aligns spoken dialogue with the corresponding
fiction text, the surrounding narrative can be used to extract
information about a character's identity, personality, age, social
role, and temporary physical or emotional state. These attributes may
be organized into natural-language voice descriptions and paired with
the character's speech. Compared with descriptions obtained only from
short audio segments, descriptions derived from longer narrative
contexts may provide a more complete characterization of the intended
voice.

This direction nevertheless requires careful distinction between
stable speaker properties and utterance-specific speaking styles.
For example, age and general personality may characterize a speaker
across multiple utterances, whereas fatigue, anger, or breathlessness
may apply only to a particular scene. Separating these two levels of
description would be important for constructing reliable voice-design
training data.

\subsection{Paralinguistic and Nonverbal Vocalization Mining}

PromptDialog may further support the extraction of paralinguistic and
nonverbal vocalizations. Existing studies, such as
NVSpeech~\citep{liao2025nvspeech}, commonly represent nonverbal sounds
using a predefined set of discrete labels. In Chinese fiction,
however, many vocal events are explicitly represented through
onomatopoeic expressions or descriptive words, including laughter,
crying, sighing, coughing, tongue clicking, and hesitation sounds.
The audio--text alignment provided by our pipeline makes it possible
to retrieve speech segments containing these events directly from
their textual occurrences.

Text-based retrieval also allows finer distinctions than a single
event label. For instance, different written forms of laughter may
correspond to different realizations, such as \textit{haha},
\textit{heh}, and \textit{hehe}. Although these expressions do not
provide exact acoustic labels, they offer a simple way to construct
more diverse subsets of nonverbal vocalizations.

As an initial analysis, we use keyword matching to identify several
common types of vocal events in the Chinese subset. Table~\ref{tab:nvs}
reports the number and total duration of the matched utterances.
Because one utterance may contain more than one keyword, the event
categories are not mutually exclusive. These statistics should
therefore be interpreted as approximate corpus-level counts rather
than verified event annotations.

\begin{table}[t]
    \centering
    \small
    \begin{tabular}{lrr}
        \toprule
        Event & Utterances & Duration (h) \\
        \midrule
        Sigh          & 114,359 & 254.18 \\
        Laughter      & 24,462  & 50.15  \\
        Crying        & 615     & 1.02   \\
        Cough  & 1,580   & 2.30   \\
        Tsk           & 452     & 0.98   \\
        \midrule
        Unique utterances
                      & 139,832 & 303.23 \\
        \bottomrule
    \end{tabular}
    \caption{Keyword-based statistics of utterances potentially
    containing paralinguistic or nonverbal vocalizations. Event
    categories may overlap because an utterance can match multiple
    keyword rules. The final row reports the number and total duration
    of unique matched utterances.}
    \label{tab:nvs}
\end{table}

Keyword matching alone does not guarantee that the corresponding sound
is acoustically present. A word may be narrated rather than performed,
and the same expression may be realized differently across contexts.
Conversely, the absence of a matching keyword does not imply that the
utterance contains no nonverbal vocalization. Narrative descriptions
such as \textit{speaking through painful sobs} or \textit{bursting into
laughter while speaking} may lead performers to insert crying,
laughter, or other vocal events that are not explicitly represented in
the dialogue text. Therefore, these automatically retrieved samples may
contain both false positives and false negatives and require additional
acoustic verification before being used as fine-grained event
annotations. Nevertheless, the results indicate that PromptDialog
contains a substantial amount of potentially useful paralinguistic
speech data.

\section{Potential Applications of PromptTTSD}
\begin{figure*}[t]
    \centering
    \includegraphics[width=\linewidth]{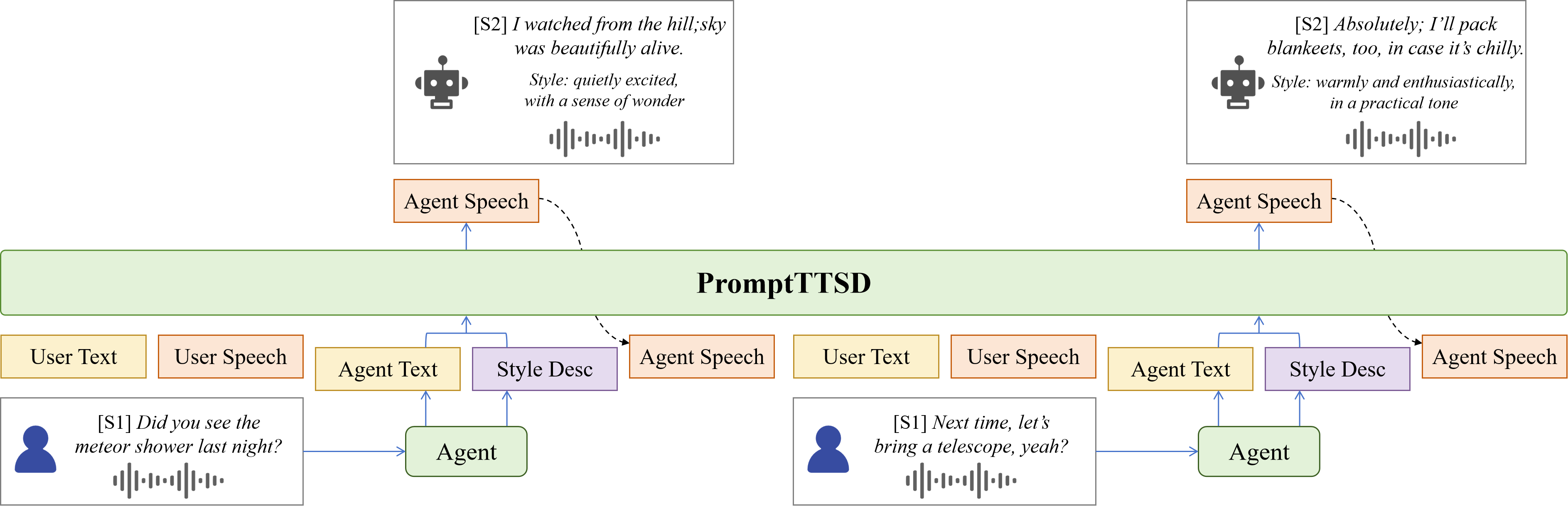}
\caption{PromptTTSD for interactive chat with open-ended style control.
Adapted from FireRedTTS-2~\citep{xie2025fireredtts}.}
    \label{fig:interactive_dialogue}
\end{figure*}

PromptTTSD is primarily designed for multi-speaker spoken dialogue
synthesis. Its direct applications include audiobook production,
character dialogue dubbing, and podcast generation. In addition, its
sentence-by-sentence generation architecture makes it naturally
compatible with interactive spoken systems. These applications have
not been systematically evaluated in this work and are discussed as
potential extensions of the model.

\subsection{Interactive Chat}

Current interactive chat systems typically rely on monologue TTS
models to synthesize each system response independently. Such models
have limited access to preceding user queries and agent responses, and
must therefore infer the appropriate emotion and prosody mainly from
the current response text. This may result in speaking styles that are
inconsistent with the dialogue context. Explicit control signals, such
as emotion labels, can partially alleviate this problem, but a small
predefined label set is insufficient to represent the diverse and
context-dependent speaking styles required in natural interaction.

As illustrated in Fig.~\ref{fig:interactive_dialogue}, the
sentence-by-sentence architecture of PromptTTSD can be naturally
extended to interactive chat. At each turn, an upstream agent generates
both the response text and an open-ended natural-language description
of its intended delivery. PromptTTSD then conditions on the response
text, the style description, and the preceding text--speech context to
synthesize the agent speech. The generated speech is subsequently
appended to the interaction history and used as context for following
turns.

Compared with emotion-label-based control, open-ended descriptions can
specify more detailed and context-dependent speaking behaviors, such
as speaking with quiet excitement and a sense of wonder, or responding
warmly and enthusiastically in a practical tone. This formulation
allows the agent to determine not only \emph{what to say}, but also
\emph{how to say it}, without restricting the intended delivery to a
fixed attribute inventory. Moreover, the decoder architecture inherited
from the base system is compatible with streaming generation, making
PromptTTSD potentially applicable to low-latency and expressive
interactive speech synthesis. We leave a systematic evaluation of
real-time interactive chat to future work.

\subsection{Action-Conditioned Speech Generation}

Interactive avatar systems typically require multiple models to
coordinate speech, facial expressions, lip movements, and body
gestures. A common approach is to drive facial and lip animation
directly from speech~\citep{fan2022faceformer,xing2023codetalker},
while related work generates upper-body or holistic co-speech motion
from the same acoustic input~\citep{qian2021speech,
zhu2023diffgesture,yi2023talkshow,chen2024diffsheg}.
This speech-driven formulation provides a practical way to synchronize
visual behaviors with spoken content in general conversational
scenarios.

However, the relationship between speech and action is not strictly
unidirectional. Actions and physical states may also constrain how an
utterance should be delivered. For example, hurried hand-waving may
suggest an anxious or urgent refusal, while speaking during running may
result in breathlessness and disrupted prosody.

Recent text-guided motion generation systems represent intended actions
using natural-language descriptions. PromptTTSD can potentially use the
same action descriptions as conditions for speech generation. For
example, a shared description such as \textit{waving his hands in a
hurry} or \textit{speaking while running and struggling to catch her
breath} may be provided to both the motion-generation model and
PromptTTSD.

Using a shared natural-language condition may reduce the coordination
gap between independently trained speech and motion models without
requiring direct feature exchange between them. This provides a
potential framework for generating more coherent speech, facial
expressions, and body movements in embodied agents and digital humans.
We leave the systematic evaluation of action-conditioned multimodal
generation to future work.

\section{Human Evaluation Protocol}
\label{app:human_evaluation}

\subsection{Style Adherence Evaluation (Style-CMOS)}

Evaluators were presented with a target style description and two
anonymized speech samples, Sample A and Sample B, in random order.
They were asked to determine which sample more closely matched the
target description in terms of emotion, prosody, speaking rate,
intensity, and manner of delivery.

Ratings were given on a five-point comparative scale. A score of
$+2$ indicates that Sample A matches the target style substantially
better than Sample B, while $+1$ indicates a slight preference for
Sample A. A score of $0$ indicates no reliable preference. Scores of
$-1$ and $-2$ indicate slight and substantial preferences for Sample B,
respectively.

\subsection{Speech Quality Evaluation (Q-MOS)}

Evaluators were presented with one synthesized speech sample and asked
to assess its overall perceptual quality, including naturalness,
intelligibility, pronunciation, fluency, prosodic continuity, and
audible artifacts.

Ratings were given on a five-point scale. A score of 1 indicates very
poor speech that is severely degraded, highly unnatural, or difficult
to understand. A score of 2 indicates poor speech with noticeable
degradation and limited usability. A score of 3 indicates fair speech
that is understandable but contains noticeable quality problems. A
score of 4 indicates good speech with only minor imperfections. A score
of 5 indicates highly natural and clear speech with no noticeable
artifacts.

\subsection{Contextual Appropriateness Evaluation (CA-MOS)}

Evaluators were presented with the dialogue context, the target
utterance, and one synthesized speech sample. They were asked to assess
whether the sample's emotion, prosody, energy, speaking rate, and manner
of delivery were appropriate for the dialogue context and the speaker's
intended state.

Ratings were given on a five-point scale. A score of 1 indicates that
the delivery strongly conflicts with the dialogue context. A score of
2 indicates major contextual mismatches. A score of 3 indicates a
generally plausible delivery with noticeable mismatches or insufficient
expressiveness. A score of 4 indicates that the delivery fits the
context well, with only minor missing nuances. A score of 5 indicates
that the delivery convincingly expresses the intended emotion and
speaking manner and is fully consistent with the context.

\section{Copyright} 

We do not redistribute the original audiobook recordings. Instead, we
provide only the corresponding YouTube links and segmentation files.
The copyright of each recording remains with its original rights
holder. The released resources are intended solely for academic
research and may not be used for commercial purposes. Upon receiving a
valid request from a rights holder or other relevant party, we will
promptly remove the corresponding links and segmentation information
from subsequent releases.

\section{Limitations}

PromptDialog currently consists predominantly of Chinese speech. The
English subset is relatively limited and is mainly used to verify that
our data-construction pipeline can be applied to English audiobooks.
We have not yet conducted a comprehensive evaluation of open-ended
style control in English. In future work, we plan to increase the
proportion of English data, retrain the model on a more balanced
multilingual corpus, and systematically evaluate its English synthesis
and style-control capabilities.

The evaluation of open-ended style control also remains limited.
Unlike predefined attributes such as emotion or speaking rate,
open-ended styles are difficult to recognize reliably using existing
automatic tools. Moreover, there is currently no established benchmark
for evaluating whether synthesized speech follows complex and
unrestricted natural-language descriptions. We therefore rely
primarily on human listening tests, which restricts the scale of the
evaluation because of their cost and time requirements. Future work
will develop a more comprehensive benchmark and evaluation protocol
covering a broader range of open-ended descriptions, speaking
behaviors, and dialogue contexts.

\bibliography{aaai2027}

\end{document}